\documentclass[conference,twocolumn,a4paper,]{IEEEtran}
\IEEEoverridecommandlockouts

\ifCLASSOPTIONcompsoc
  \usepackage[nocompress]{cite}
\else
  \usepackage{cite}
\fi

\ifCLASSINFOpdf
\else
\fi

\usepackage{xifthen}
\usepackage{xcolor}
\usepackage{mathtools}
\usepackage{amsfonts}
\usepackage{nicefrac}
\usepackage{booktabs}
\usepackage{semantic}
\usepackage{url}
\usepackage{hyperref}
\usepackage{tikz}
\usepackage{svg}
\usepackage[mathlines,running]{lineno}
\usepackage{relsize}

\usetikzlibrary{calc}
\usetikzlibrary{math}

\usepackage{stackengine}
\usepackage{oplotsymbl}
\usepackage[most]{tcolorbox}
\newcommand{\zenodo}{\url{https://doi.org/10.5281/zenodo.14168707}}

\newcommand{\codestring}[1]{\lstinline{"#1"}}

\newtcbox{\quotescitation}{%
  enhanced,
  nobeforeafter,
  colframe=white,
  colback=blue!100!red!70!green!35,
  size=small,
  boxrule=0pt,
  fontupper=\scriptsize\sf,
}

\newtcolorbox{quotes}[1][]{
  enhanced,
  colback=black!5!white,
  colframe=blue!100!red!70!green!35,
  sharp corners,
  toprule=0mm,
  bottomrule=0mm,
  rightrule=0mm,
  leftrule=.4em, %
  overlay={
    \node[anchor=south east,xshift=.1em,yshift=-.7em] at (frame.south east) {\quotescitation{#1}};
  },
}

\newcommand{\mn}[1]{\mathrm{#1}}
\newcommand{\ms}[1]{\mathsf{#1}}

\newcommand{\pbr}[1]{{\left( #1 \right)}}
\newcommand{\sbr}[1]{\left[ #1 \right]}

\newcommand{\at}[1]{{\left[ #1 \right]}}
\newcommand{\of}[1]{{\left( #1 \right)}}

\newcommand{\pair}[2]{\left ( {#1}, {#2} \right )}

\newcommand{\set}[1]{\left \{ #1 \right \}}
\newcommand{\bitset}[1][]{\set{0,1}\ifthenelse{\isempty{#1}}{}{^{#1}}}
\newcommand{\natset}{\mathbb{N}}
\newcommand{\intset}{\mathbb{Z}}

\newcommand{\groupset}{\mathbb{G}}
\newcommand{\obseq}{\approx}

\newcommand{\stateop}{\mn{state}}
\newcommand{\eventop}{\mn{event}}

\newcommand{\macop}{\mn{mac}}

\newcommand{\rf}{\circletfill}

\newcommand{\rh}{\circletfillhb}
\newcommand{\rc}{\circlet}

\usepackage{soul}
\usepackage{bbding}

\definecolor{highlightcolour}{rgb}{.8,1,.5} %
\definecolor{highlightcolour}{rgb}{.5,.7,1} %
\definecolor{highlightcolour}{rgb}{.9,.8,1} %
\definecolor{highlightcolour}{rgb}{.5,.95,1} %
\definecolor{highlightcolour}{rgb}{.8,.8,.8} %
\definecolor{highlightcolour}{rgb}{1,.95,.5} %
\sethlcolor{highlightcolour}

\newcommand{\sidenote}[1]{\todo[size=\scriptsize,color=black!20!blue!10!white,bordercolor=black!20]{#1}}
\newcommand{\sn}[2][]{{\ifthenelse{\isempty{#1}}{\sidenote{#2}}{\sidenote{{\bf #1:} #2}}}}
\newcommand{\hsn}[3][]{{\setulcolor{red}\ul{#3}}{\ifthenelse{\isempty{#1}}{\sidenote{#2}}{\sidenote{{\bf #1:} #2}}}}
\newcommand{\ilnote}[1]{\todo[inline,size=\small,noshadow,color=highlightcolour!30!white,bordercolor=highlightcolour,caption={}]{{\hfill\Large\color{highlightcolour!80!black}\PencilLeftDown\\\vspace{-1em}}\begin{minipage}{.95\columnwidth}#1\end{minipage}}}
\newcommand{\tn}[2][]{{\ifthenelse{\isempty{#1}}{\ilnote{#2}}{\ilnote{#2{\newline\hfill\sc \color{highlightcolour!40!black}-- #1\;}}}}}
\newcommand{\htn}[3][]{{\hl{#3}}{\ifthenelse{\isempty{#1}}{\ilnote{#2}}{\ilnote{#2{\newline\hfill\sc \color{highlightcolour!40!black}-- #1\;}}}}}

\newcommand{\procstop}{\mathbf{0}}
\newcommand{\pinonce}[1]{\nu #1.}
\newcommand{\piread}[2][c]{\ms{in}\of{#1, #2}}
\newcommand{\piwrite}[2][c]{\ms{out}\of{#1, #2}}

\newcommand{\pilet}[2]{\ms{let}\ {#1}\ \ms{in}\ {#2}} %

\newcommand{\piinsert}[2]{\ms{insert}\ #1\of{#2}}
\newcommand{\piget}[3]{\ms{get}\ #1\of{#2}\ \ms{in}\ {#3}} %
\newcommand{\pigetse}[5]{\ms{get}\ #1\of{#2}\ifthenelse{\isempty{#3}}{}{\ \ms{suchthat}\ {#3}}\ \ms{in}\ {#4}\ifthenelse{\isempty{#5}}{}{\ \ms{else}\ {#5}}} %

\newcommand{\traceevent}[1]{e_{\textsc{#1}}}
 
\usepackage{listings}
\usepackage{algorithm}
\usepackage[noend]{algpseudocode}
\usepackage{algorithmicx}
\lstdefinelanguage{None}
{
  basicstyle=\ttfamily\small,
}

\usepackage[]{listings}

\definecolor{keywordcolor}{RGB}{171, 64, 56}
\definecolor{morecomment}{RGB}{0,128,0}
\definecolor{stringcolor}{RGB}{255,0,0}
\definecolor{equalscolor}{RGB}{102,0,102} %
\definecolor{coloncolor}{RGB}{102,0,102}  %
\definecolor{selfcolor}{RGB}{0,102,51}

\lstdefinelanguage{AsmetaL} %
{   keywords={module, par, endpar, if, endif, then, else, seq, endseq, 
        signature, definitions, asm, import, function, domain, main, rule, macro, 
        invariant, over, choose, let, endlet, with, ifnone, forall, abstract, default, 
        init, do, agent, dynamic, controlled, monitored, in, out, static, derived, extend, 
        subsetof, switch, case, endswitch, enum, CTLSPEC, LTLSPEC, JUSTICE, step, set, scenario, load},
	sensitive=true,
    morecomment=[l]{//}, 
    morecomment=[s]{/*}{*/},
	morecomment=[l][\color{white}\tiny]{'},%
	morestring=[b]",
    tabsize=1, 
    columns=fullflexible, 
    basicstyle=\scriptsize\sffamily, 
    captionpos=b,
    escapeinside={(*}{*)},
    numbers=left,
    numberstyle=\scriptsize, %
    numbersep=5pt, %
    xleftmargin=10pt %
    }
\usepackage{wrapfig}
\usepackage{url}
 
\usepackage{soul}
\usepackage{makecell}
\usepackage{arydshln}
\newlength{\protocolArrowLength}
\newlength{\partysize}
\newcolumntype{B}[1]{>{\centering\arraybackslash}b{#1}}
\newcolumntype{C}[1]{>{\centering\let\newline\\\arraybackslash\hspace{0pt}}m{#1}}

\newcommand{\passwordelement}{\mathrm{PE}}
\newcommand{\hash}{\mathrm{H}}
\newcommand{\keyderivationfunction}{\mathrm{KDF}}

\newcommand{\PE}{\passwordelement}
\newcommand{\KDF}{\keyderivationfunction}

\newcommand{\cn}{\mathrm{CN}} %
 
\usepackage[normalem]{ulem} %

\usepackage[inline]{enumitem}

\newcommand{\tokens}[1]{\texttt{#1}}

\usepackage{multirow}

\usepackage{xspace}
\newcommand{\ttt}[1]{%
  \begingroup
  \ttfamily
  \hyphenchar\font=45
  #1%
  \endgroup
}

\def\BibTeX{{\rm B\kern-.05em{\sc i\kern-.025em b}\kern-.08em
    T\kern-.1667em\lower.7ex\hbox{E}\kern-.125emX}}
\begin{document}
\title{What a Mesh: Formal Security Analysis of WPA3 SAE Wireless Authentication}

\author{\IEEEauthorblockN{Roberto Metere}
\IEEEauthorblockA{\textit{University of York}\\
York, UK \\
roberto.metere@york.ac.uk}
\and
\IEEEauthorblockN{Mario Lilli}
\IEEEauthorblockA{\textit{King's College London}\\
London, UK \\
mario.lilli@kcl.ac.uk}
\and
\IEEEauthorblockN{Luca Arnaboldi}
\IEEEauthorblockA{\textit{University of Birmingham}\\
Birmingham, UK \\
l.arnaboldi@bham.ac.uk}
\and
\IEEEauthorblockN{Elvinia Riccobene}
\IEEEauthorblockA{\textit{University of Milan}\\
Milan, Italy \\
elvinia.riccobene@unimi.it}
}

\maketitle

\begin{abstract}
The latest Wi-Fi security standard, IEEE 802.11, includes a secure authentication protocol called SAE, whose use is mandatory for WPA3-Personal networks.
The protocol is specified at two separate but linked levels: a traditional cryptographic description of the communication logic between network devices, and a state machine description that realises the former in each single device.
Current formal verification efforts focus mainly on communication logic.
We present detailed formal models of the protocol at both levels, provide precise specifications of its security properties, and analyse machine-checked proofs in ProVerif and ASMETA.
The integrated analysis of the above two models is particularly novel, enabling us to identify and address several issues in the current IEEE 802.11 specification more thoroughly than would have been possible otherwise, leading to several official revisions of the standard.
\end{abstract}

\IEEEpeerreviewmaketitle

\section{Introduction}
\label{sec:intro}

For years, security experts have undertaken the challenging work of transforming ambiguous natural language specifications into precise mathematical models~\cite{cremers2017comprehensive,cremers2020formal,arnaboldi2019formal,metere2017automated,ye2024user} allowing holistic analysis, much beyond security testing.
Research efforts from the formal methods community have developed increasingly sophisticated tools for automating formal security verification ~\cite{basin2022tamarin,blanchet2021proverif,armando2005avispa,arnaboldi2019poster,metere2022automating}.
Formal verification proves worthwhile precisely because it catches entire classes of vulnerabilities that emerge from specification flaws that may persist across all implementations regardless of code quality.

Introduced in 2018, WPA3 was designed to improve upon the security guarantees of WPA2-based Wi-Fi networks.
However, recent work has shown that WPA3 remains susceptible to previously identified vulnerabilities in certain implementations~\cite{vanhoef-sp2020-dragonblood}.
This work complements prior research by presenting the first formal verification of WPA3’s authentication protocol, SAE (Simultaneous Authentication of Equals), and providing a rigorous security analysis.

We worked on the WPA3-SAE security protocol as specified in the IEEE 801.11 standard~\cite[§12.4]{ieee802.11:2020}.
The SAE protocol was already included in IEEE 802.11-2016, but WPA3 made it mandatory for WPA3-Personal networks only in the 2020 revision.
Its specification is divided into two separate but linked parts that specify the protocol at two different levels.
The first part, subsections §12.4.1-7, provides a cryptographic description of the communication logic between two parties, which we call {\em Communication level}.
This main, traditional specification is complemented by a second part, subsection §12.4.8, that \emph{instantiates} the protocol specified in the first section as a finite automaton running in a single device, which we call {\em Device level}.

Our research uses two verification approaches: the traditional Dolev-Yao symbolic analysis at the Communication level, and a transition system analysis at the Device level.
The two specifications are largely self-contained (even when linked), so we modelled them independently.
This allowed us to identify ambiguities or weaknesses in the specification, leading to a better understanding of the origins of some recent (now patched) implementation flaws~\cite{vanhoef-sp2020-dragonblood}, as well as a few potential vulnerabilities in software that complies with the specification.
For some properties, insights gained from one model informed the analysis of the other, improving overall accuracy.
This cross-model combination is informal: rather than being guided by counterexamples or failed properties, it relies on propagating missing or ambiguous controls between levels, thereby exposing ambiguities and contradictions in the specification itself.
Our formal analysis identified previously unrecognised vulnerabilities in the WPA3 specification and enabled us to formally demonstrate their existence.
We used the resulting proofs to design and justify concrete patches to the standard\footnote{Hyperlinks and references to documents are omitted to maintain anonymity; a redacted version is provided in the Appendix.}, the majority of which have since been accepted and integrated into the latest revision of IEEE 802.11~\cite{ieee802.11:2024}.
Each model captures complementary aspects and properties of the protocol, making our analysis more comprehensive than traditional single-model verification, and provided new insights that significantly improved our understanding and patching of the standard.

Some of our findings can be linked to previous attacks on WPA3, suggesting their effectiveness in identifying and describing new vulnerabilities.
For example, the reflection attack we describe in Section~\ref{sec:replay-attack} was known since 2015 but affected several implementations up to 2019.
Another example is the memory exhaustion attacks described by Vanhoef and Ronen in ~\cite{vanhoef-sp2020-dragonblood}, which share the same nature of deadlocks that we have noticed at the Device level in Section~\ref{sec:findings}.
Importantly, our analysis was conducted on a version of the IEEE 802.11 standard that already incorporated the patches proposed in~\cite{vanhoef-sp2020-dragonblood}. 
Despite this, our models still expose security-relevant deadlocks and vulnerabilities, indicating that these fixes were insufficient at the specification level.
Our findings contribute to the development of more secure wireless protocols.

We summarize our contributions as follows.

\begin{enumerate}
  \item We propose the first formal analysis of the SAE protocol, modelled at both its \textit{communication} and \textit{device} levels.
  As part of this process, we added equations to our analysis that capture the modular division operations involving the group exponentiation (Section~\ref{sec:division}), which are general and reusable in other contexts where groups are based on finite field or elliptic curve cryptography.
   
  \item We report substantial results from our security analysis of the latest WPA3-SAE specification. 
  We identify over 20 previously undocumented issues, including weaknesses, ambiguities, missing guidelines, contradictions, and security-relevant flaws (see Appendix~B).
  Not all of these issues directly yield practical attacks; however, each represents a previously undocumented flaw in the specification that can affect compliant implementations.
  Section~\ref{sec:findings} presents a representative subset with clear security implications.
  The remaining findings include corrections to behaviours that should have been explicitly forbidden, resolution of unspecified password-identifier logic, clarification of contradictory requirements, and completion of missing protocol elements.
 
  \item Finally, we propose formally verified solutions for the identified weaknesses; these solutions have since been incorporated into the standard.
  We make our models and results available at \zenodo{}.
\end{enumerate}

\textit{The rest of the paper is structured as follows:}
Section~\ref{sec:related-works} summarises work in this area and related literature.
In Section~\ref{sec:wpa3sae}, we introduce the Simultaneous Exchange of Equals (SAE).
We then explain our verification process in Section~\ref{sec:methodology}, provide details of our analysis of the WPA3-SAE formal models in Section~\ref{sec:formal-models}, and present our results in Section~\ref{sec:findings}.
We finally discuss our findings in Section~\ref{sec:real-world} in relation to real-world WPA3 implementations and draw our conclusions in Section~\ref{sec:conclusions}.
\section{Related Works}
\label{sec:related-works}

The use of multiple tools for the verification of protocols has come under substantial attention recently~\cite{arquint2023generic,morio2020modular}, but this is not altogether new as it was one of the core principles of AVANTSSAR~\cite{armandoAVANTSSAR} a decade ago, whose intermediate language is based on the \textsf{AnB} logic, and could not be used to capture enough security aspects of complex and modern protocols.
Early efforts to combine analyses across different verification tools date back to 2019~\cite{arnaboldi2019poster}; follow-up work has since improved the reliability and theoretical foundations of multi-tool analyses~\cite{cheval2022sapic}.
In contrast, our focus is not on tool integration, but on applying distinct tools, such as ProVerif and ASMETA, to explore complementary aspects of the same protocol.

Successively, an independent work extended the security guarantees typically associated with protocol designs to their actual implementations~\cite{morio2020modular}.
This is achieved by instrumenting common cryptographic libraries and network interfaces with a runtime monitor.
We sacrifice the ability to directly refer to runtime implementations to remain in the formal domain and ensure full verification coverage at both analysed levels.

An interesting recent technique presents a methodology that encapsulates global traces and uses invariants over local memory snapshots to achieve modular verification of protocol implementations~\cite{arquint2023generic}.
Notably, this approach is well-suited for direct application to existing codebases like WireGuard, whereas our focus lies elsewhere.

While our approach is conceptually related to counterexample-guided abstraction refinement (CEGAR)~\cite{clarke2003counterexample}, it differs fundamentally.
Rather than employing a formal refinement loop over a single abstraction, we use complementary abstractions to provide orthogonal views of SAE behaviour.
This enables broader coverage of security properties and more direct support for patching the specification, rather than refining artefacts introduced by a particular abstraction.

Probably the closest work to ours can be found in an early verification of TLS1.2~\cite{bhargavan2013implementing}, where authors identified that the state machine is an important part of the implementation and performed some verification on it.
Their verification process is limited to type-checking the state machine of the TLS protocol.
Type checking can verify that the program adheres to the specified types, which in this context relate to the correct sequence of operations and data handling in the protocol.
Although this is undoubtedly a first step, authors have never linked the state machine to the upper layers of the protocol, as we do in our analysis (as we model the same layer).
Additionally, we verify properties beyond type checking.

We note that formal analysis is not the only way to analyse a protocol, and indeed, authors have already run security evaluations of similar protocols~\cite{vanhoef-sp2020-dragonblood,vanhoef-ccs2017}, i.e., the DragonFly protocol, which is the predecessor of the SAE key exchange.
Their work identified several vulnerabilities, focusing on side-channel attacks~\cite{vanhoef-sp2020-dragonblood} and communication-layer attacks~\cite{vanhoef-ccs2017} in several real-world devices that implemented the key exchange.
Their work contributed to the 2020 version of the IEEE 802.11 standard that we analysed; we find our work complementary to theirs, as we add formal methods to the analysis.
\section{WPA3-SAE}
\label{sec:wpa3sae}

WPA3 is secured using the Simultaneous Authentication of Equals (SAE) protocol, which has been introduced to replace the Pre-Shared Key (PSK) exchange used in WPA2, known to be vulnerable to attacks~\cite{vanhoef-ccs2017}.
SAE provides authentication at the Link layer, which is part of the TCP/IP stack.
It was originally introduced in 2016 as part of IEEE 802.11 - IEEE Standard for Information Technology--Telecommunications and Information Exchange between Systems~\cite{ieee2007ieee}, with claims to resolve the previous vulnerabilities affecting PSK and WPA2.
We focus our security analysis on the protocol version of 2020, following the IEEE 802.11w~\cite{ieee802.11:2020} specification.

\subsection{The SAE Key Exchange}
\label{sec:sae-key-exchange}

The SAE key exchange is a two-party protocol that takes two rounds of messages.
Each round is symmetric so that we do not have a notion of an Initiator and Responder or of a Supplicant and Authenticator.
Each side may initiate the protocol simultaneously, such that each side views itself as the {\em initiator} for a particular run of the protocol.
This design is necessary to address the unique nature of the mesh basic service set~\cite{ieee802.11:2020}.
Several variants of SAE are specified; nonetheless, we only focus on the variant of SAE adopted in the WPA3 protocol, especially the variant that operates in finite field cryptography\footnote{Our models abstract the mathematical operations in the finite field, so that they are analogous to the Elliptic Curve specifications.}.

\begin{figure}[!ht]
  \caption{SAE protocol in WPA3~\cite[§12.4]{ieee802.11:2020} as specified in the standard.
  $A$ and $B$ share the secret password $w$ and computed $\PE$ in private; $\PE \in \groupset$ and $\groupset$ is a subgroup of $\intset^{\star}_p$ of order $q$; $i$ is a counter.
  $B$ is symmetric to $A$ and thus omitted for brevity.}
  \label{fig:sae-communication}
  \centering\includegraphics[width=.9\columnwidth]{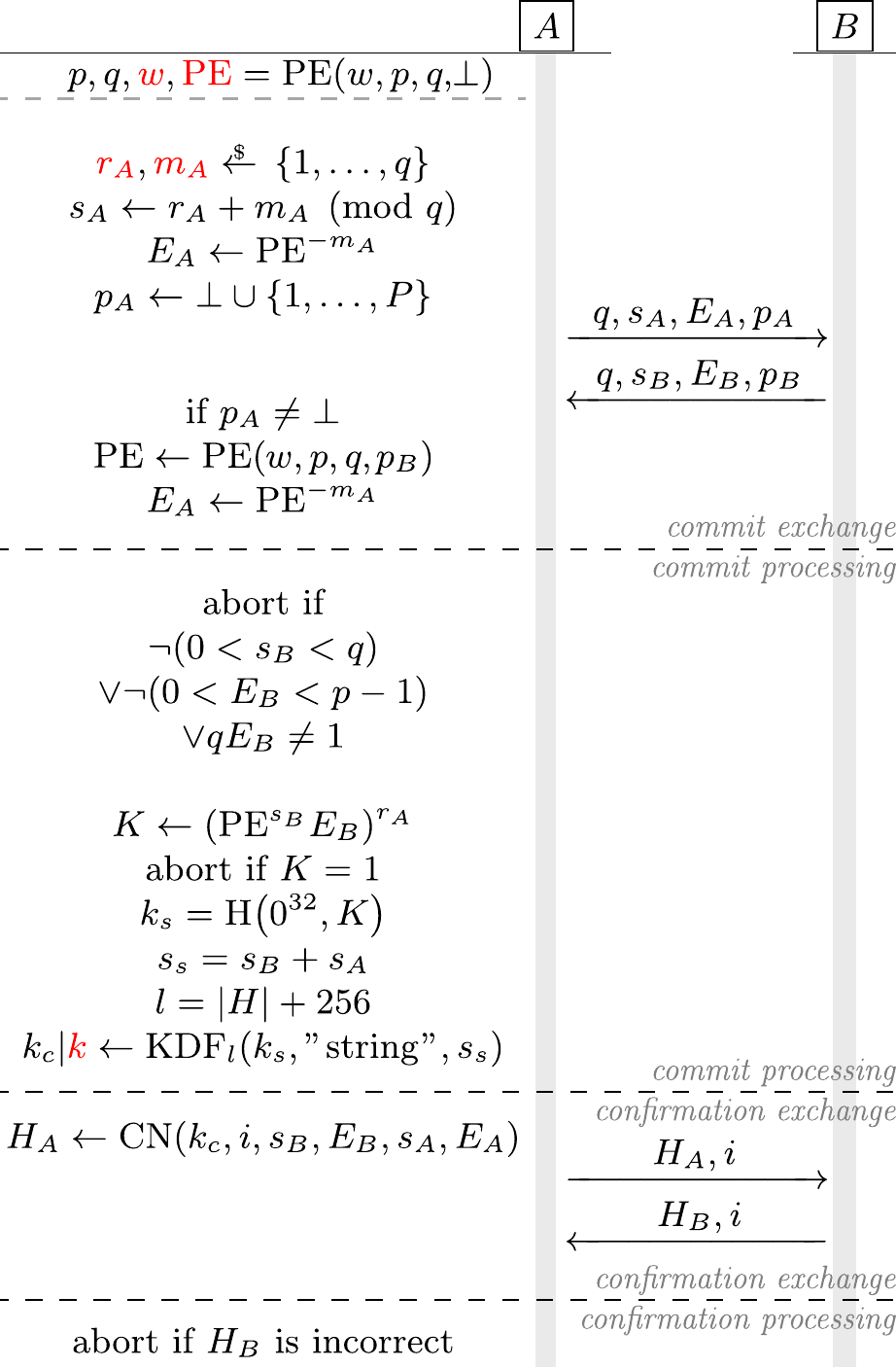}
\end{figure}

The SAE protocol operates in a finite cyclic group, which may be based on either a finite field or on elliptic curve cryptography.
We adopt the former's notation for simplicity, but it can be generally extended to the latter.
Formally, we denote $\groupset$ as a common and public subgroup of $\intset^{\star}_p$ of multiplicative order $q$, where $q$ is a Sophie Germain prime, i.e., $p = 2q + 1$,  where the discrete logarithm problem is assumed to be hard.
The protocol also uses $\hash$ to represent a hash-based message authentication code (HMAC).
Two remote parties, Alice and Bob, share a common secret password from which they apply a transformation to calculate a corresponding {\em password element}, $\PE$.
Using $\PE$, a secret element can be derived as described by the Key Derivation Function $\KDF$, defined in the standard~\cite{ieee802.11:2020}: $\PE$ is calculated as a function of the password, the agreed group parameters, and, optionally, a password identifier.
The (last) confirmation phase also specifies a confirmation function $\cn$, which is also defined in the standard.
Both $\KDF$ and $\cn$ are calls to the hash function $\hash$.
The SAE protocol runs in two rounds: the {\em commit exchange} and the {\em confirmation exchange}, as illustrated in Figure~\ref{fig:sae-communication} for the peer $A$, which communicates with a symmetric peer $B$.

\subsection{The SAE Message Handling}
\label{sec:sae-realisation}

The protocol is instantiated by the finite-state machine described in §12.4.8 of the standard; Figure~\ref{fig:sae-device} reports a simplified state/transition diagram of each Protocol Instance (PI).
Message handling is described through interactions among (concurrent) PIs and two other entities: Station Management Entity (SME), and Parent Process (PP).

\subsubsection*{SME}
This entity is a component responsible for managing various aspects of a wireless station or device.
It primarily concerns the management of the physical and medium access control (MAC) layers of the 802.11 protocol.
    SME provides an interface for higher-layer protocols to interact with the lower layers and handles tasks related to the wireless station's configuration, operation, and maintenance.
However, this entity is not strictly described in the standard~\cite{ieee802.11:2020}, which states:
\begin{quotes}[§6.1, pg. 314]
   Some functions of the SME are specified in this standard.
\end{quotes}
Moreover, the description is fragmented and given incrementally in different parts of the documentation.
This easily causes misinterpretation and misunderstanding.
 
\subsubsection*{PP}
The parent process manages the database of the protocol instances (PIs).
It performs several tasks, such as allocating and deallocating instances, and keeping track of their respective states.
Additionally, it is responsible for routing incoming messages from the environment to the correct PI (allocating a new instance when needed).
To accomplish these operations, it sends a suitable event to trigger the PI's state change.
It also keeps the database up to date based on events received from both the SME and the PIs.
\subsubsection*{PIs}
The processes that ultimately realise the protocol enact the behaviour specified as a state machine. %
State changes are triggered by messages and events received from the PP, but transitions fire during message content analysis.
Transition execution is accompanied by actions that the PI executes before entering a state.
Such actions consist of messages for the peer or generation of \textit{error} or \textit{completion} output events to be sent to the PP for subsequent deallocation of the instance.

\begin{figure}[h!]
  \caption{State machine instantiating the SAE protocol in WPA3~\cite[§12.4.8]{ieee802.11:2020}.}
  \label{fig:sae-device}
  \centering\includegraphics[width=\columnwidth]{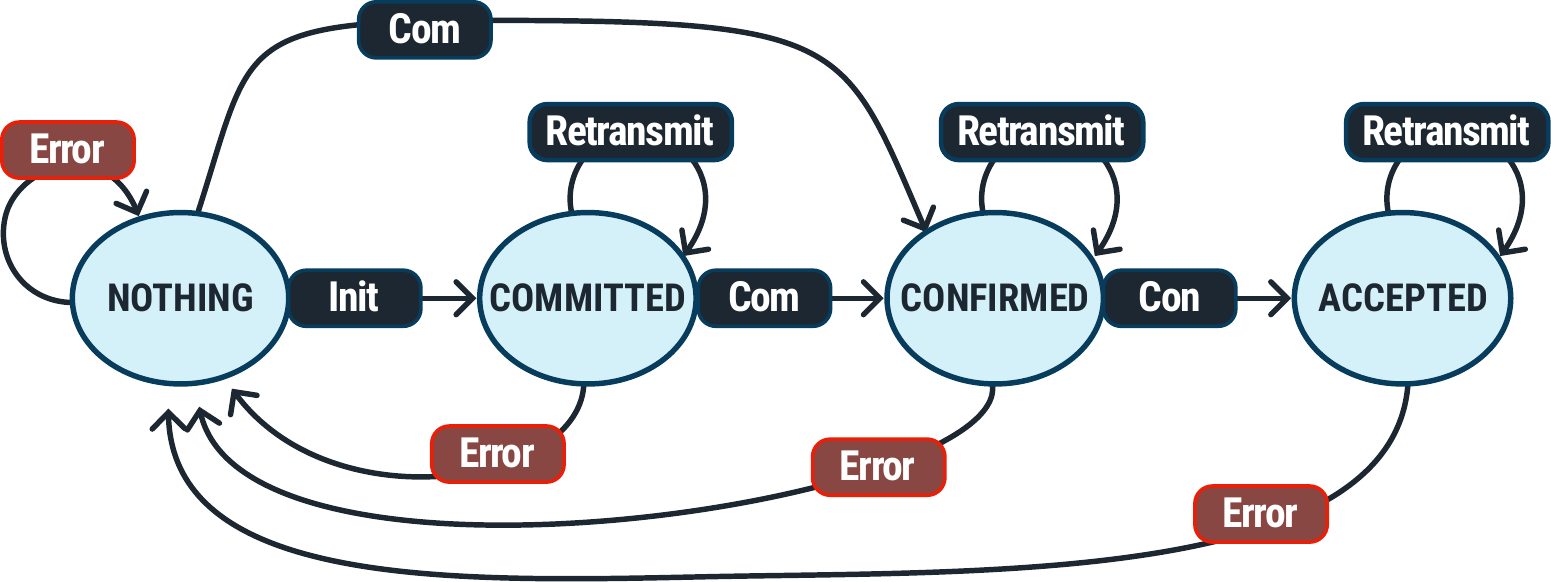}
\end{figure}

Fig.~\ref{fig:sae-device} depicts a simplified version of a PI's state machine consisting of four states and having transitions labelled with the relevant events sent by the PP and able to trigger a state transition (note that we omit information regarding the received messages that might prevent transition firing, and actions performed by the PI when a transition fires).

A PI is in the state
 \textsf{Nothing} initially, as a new instance, and finally, as a terminal state before being deallocated in case of error. 
 Depending on their allocation, new  PI  immediately transitions out of the Nothing state to either \textsf{Committed} or \textsf{Confirmed}. Protocol instances that transition into the \textsf{Nothing} state shall immediately and irretrievably be deleted\footnote{This relevant information is reported only in section 12.4.8.2.2 of \cite{ieee802.11:2020}}. 
 A PI enters state \textsf{Committed} when it receives the event \textsf{Init} from the PP and has sent an SAE \emph{commit} message to the peer. In this state, the PI waits for the \textsf{Com} event from the PP and moves to state \textsf{Confirmed}, having sent an SAE  \emph{confirm} when it receives the correct SAE \emph{commit} message from the peer. The PI remains in this state if a \textsf{Retransmission}\footnote{This event groups several wrong (but manageable) error cases that can happen, e.g., wrong order of the received messages, wrong received group, etc.} event happens. When an unmanageable error occurs, a \textsf{Del} event is sent to the PP, and the PI moves to state Nothing to be deallocated. 
In the state \textsf{Confirmed}, the PI operates as in state Committed in case of events \textsf{Del} and \textsf{Retransmission}, and moves to state \textsf{Accepted} on the event \textsf{Con} from the PP and by receiving a peer's correct \textit{commit} message. 
In the (final) \textsf{Accepted} state, events \textsf{Retransmission} and \textsf{Del}, if they happen, are dealt with as in the previous states; otherwise, the PI has successfully concluded the process.

\section{Security Analysis}
\label{sec:methodology}

The main contributions of our work are twofold: (i) the first formal analysis of the WPA3-SAE and the verification of its security properties, and (ii) validated patches to the IEEE 802.11 specification.

The approach we used for modelling and verifying the WPA3-SAE protocol reflects the separation of the protocol specification into two linked sections: \textit{one}
describing the cryptographic protocol of the communication logic between two parties (Communication level), and \textit{another} that implements the first description realizing it as a finite automaton running in a single device (Device level).
Therefore, we proceeded by specifying formal models separately at either level.

The formalism at the Communication level uses a well-established modelling technique, whose analysis is based on the $\pi$-calculus as specification formalism.
Two tools are state-of-the-art for processing $\pi$-calculus mechanized models, Tamarin~\cite{cheval2022sapic} and ProVerif~\cite{blanchet2021proverif}.
Given that SAE specifications are based on group theory, Tamarin was our first choice for its algebraic theories that natively consider associativity and inverses.
However, SAPIC+ (which handles pi-calculus in Tamarin) had insurmountable limitations in parsing the model.
At that point, either we could renounce using pi-calculus in favour of multiset-based rules, or we could switch to ProVerif, assuming we could include enough algebra, which we did.
In comparison, one of our limitations is that we miss some of Tamarin's built-in associative equations.

The formalism at the Device level is based on state machine modelling, whose verification is mechanized and runs on the state-of-the-art formal framework, ASMETA~\cite{bombarda2024asmeta}.
It uses state-based transition systems as its specification formalism (specifically, Abstract State Machines~\cite{ASMbook,bookModelingCompanionASM2018}), which are efficiently supported through the tool-set ASMETA.
ASMETA has been used for swift model editing (with AsmetaL notation), validation (with AsmetaV~\cite{carioni2008scenario} and its notation Avalla), and verification using temporal-logic properties via model checking (with AsmetaSMV, which maps AsmetaL models to the model checker NuSMV~\cite{cimatti1999nusmv}).

As a result, we could leverage the strengths of each model to cover a comprehensive range of aspects and verification goals that are contextually relevant to the protocol specification (\textit{independent analysis}).
For example, the ProVerif model is unsuitable for modelling threats to availability properties (e.g., deadlocks that can cause denial-of-service), which are instead modelled in ASMETA.
For some properties, results in one model informally hinted the other in capturing specification aspects that were misplaced (e.g., reflection attacks).
This approach allows for a generally deeper study of a protocol's security than would otherwise be possible.
Examples are
(i) deadlock-inducing conditions and incomplete state-recovery logic, which were hardly visible in isolation;
(ii) implicit logics at the Device level diverged from specification at the Communication level; or
(iii) the weaknesses in session validation, at the Communication level, which led to further constraints and condition checks that we manually crafted in the state machine verification (Device level).

\subsection{Trust model}
\label{sec:trust-model}
We adopt the standard symbolic trust model for WPA3-SAE, assuming an active Dolev-Yao adversary~\cite{dolev-yao} with full control of the communication channel, including message interception, replay, injection, and impersonation.
Cryptographic primitives, including hash functions and group operations, are treated as ideal and unbreakable, with no algebraic leakage beyond their specified interfaces.
Fresh randomness is assumed for all ephemeral values.
The adversary is computationally unbounded but cannot violate cryptographic assumptions and is restricted to online password guessing only, ruling out offline dictionary attacks by construction.

The Dolev-Yao model can only test for SAE's fundamental cryptographic guarantees, so we coupled ProVerif with ASMETA to also capture state-machine level properties, and model protocol state explicitly.
As state transitions and ephemeral secrets are concrete we included race conditions, retries, and could spot several subtle implementation flaws.
The supplicant and authenticator are modelled as honest principals that correctly implement the SAE state machine and initially share a low-entropy password, assumed uncompromised at session start.

\subsection{Security Properties}
\label{sec:methodology-security-analysis}

Security properties analysed at either the Communication level or the Device level reflect typical verification goals for which automated tools are suitable when parsing their models.

\subsubsection{Communication level}
\label{sec:methodology-security-analysis-protocol}
we analyse properties common to many other key exchange protocols.
We employ an abstraction whose certification implements the Dolev-Yao attacker, so that violations of these properties reveal potential attack paths.
The properties we verify in the $\pi$-calculus
are commonly desirable to other similar protocols~\cite{hao2018analyzing}:
(i) \textbf{correctness} where parties can finalize the exchange and share the same key,
(ii) \textbf{authentication} where messages are sent from legitimate entities,
(iii) \textbf{secrecy} of the shared session key and the password element, and
(iv) \textbf{perfect forward secrecy} where leaking of keys does not compromise future keys.

\subsubsection{Device level}
\label{sec:methodology-security-analysis-state-machine}
we focus on the compartmentalised analysis of individual agents - PIs, a PP, and an SME - each with its own Abstract State Machine (ASM) that communicates with the others via events.
At this level, we concentrate on the operational correctness of message handling, synchronisation mechanisms, queue implementations, and state reachability within the protocol.
We note that at this level, the exploration leads to the discovery of possible weaknesses, rather than a specific attack, as the attacker is not modelled.
These weaknesses are then studied manually to evaluate if they can be exploited.
Since security and authentication properties have been handled by verification at the Communication level, where the messages' content is thoroughly checked, during our analysis of the agents' machines, we focused on the following verifying properties, listed along with their corresponding CTL formulas:
(i) \textbf{safety} where bad configurations do not happen,
(ii) \textbf{reachability} of desired configurations, and
(iii) \textbf{deadlock-free} where it is always possible to exit from any given configuration.
Further CTL formulas have been stated to evaluate corner cases, such as checking the \textit{edge safety} property to ensure agents do not miss events due to a lack of RAM or to simultaneous reception of events from multiple agents.
At the Device level, scenario-based simulations using the AsmetaV tool provide an additional layer of assurance by validating the model's compliance with its specifications.

\subsection{Validated Solutions}

The final stage was to validate the results and patch the attacks.
This effort was twofold:
(i) formally verify that our proposed solutions fixed the vulnerabilities by running the tool on updated models; and
(ii) coordinate with the IEEE 802.11 Working Group to update the standard to reflect the new verified patches.

Regarding the second stage, we were added as designated experts to some IEEE802.11p Working Group meetings, where maintainers of the 802.11 specification regularly meet with developers, engineers, and managers from vendors and manufacturers.
Most of the attendees, including Dan Harkins\footnote{The inventor of SAE~\cite{harkins2008simultaneous}.}, were involved in vetting our recommended changes for the next revision of the standard.
After explaining our points in several meetings and attending panel meetings, almost all our corrections were accepted and introduced into the standard (19/25 accepted)~\footnote{REDACTED FOR ANONYMITY, we provide the anonymised working group minutes in our artefact: \zenodo{}, and will include them in the appendix of the paper.}.
Remarkably, although six proposed changes were not accepted, they were acknowledged as distinct issues, but deemed {\em too unlikely} to warrant a re-specification.
This highlights a key difference between formal analysis and standardisation practice: in the formal setting, any potential threat constitutes an error, whereas in practice, considerations such as performance, integration, and usability may outweigh risks deemed highly unlikely.
As a result, some of our findings would require stronger empirical support or practical validation to motivate re-specification, which we leave as a limitation of this work.

\section{Formal Models of WPA3-SAE}
\label{sec:formal-models}

The first step of our analysis required us to model the Communication and Device levels of the protocol separately.

\subsection{Symbolic Models of Communication Level}
\label{sec:pi-models}
As shown, the $\pi$-calculus code within the boxes in Fig.~\ref{fig:pi-model-vanilla} models the protocol scheme presented in Fig.~\ref{fig:sae-communication}.

\begin{figure}[h!]
\caption{The vanilla description of the SAE protocol.}
    \label{fig:pi-model-vanilla}
\small
\begin{tikzpicture}
  \node[anchor=south west,inner sep=0] at (0,0) {
  \begin{minipage}{.5\columnwidth}
    \begin{equation*}
      \begin{array}{r|rl}
           \color{black!50}\mathsmaller{\mathsmaller{ 1}} & \,P_L <-\ & \piget{t_p}{=\!L, =\!R, =\!pw, \PE}{}
        \\ \color{black!50}\mathsmaller{\mathsmaller{ 2}} &           & \pinonce{r_L}\pinonce{m_L}
        \\ \color{black!50}\mathsmaller{\mathsmaller{ 3}} &           & \pilet{s_L = r_L + m_L}{}
        \\ \color{black!50}\mathsmaller{\mathsmaller{ 4}} &           & \pilet{E_L = \PE^{-m_L}}{}
        \\ \color{black!50}\mathsmaller{\mathsmaller{ 5}} &           & \pilet{c_L = \pair{s_L}{E_L}}{}
        \\ \color{black!50}\mathsmaller{\mathsmaller{ 6}} &           & \piwrite{c_L}; \piread{c_R};
        \\ \color{black!50}\mathsmaller{\mathsmaller{ 7}} &           & \pilet{s_R, E_R = c_R};
        \\ \color{black!50}\mathsmaller{\mathsmaller{ 8}} &           & \pilet{K = \left( \PE^{s_R}E_R \right)^{r_L}}{}
        \\ \color{black!50}\mathsmaller{\mathsmaller{ 9}} &           & \pilet{k_s = \hash\of{0^{32}, K}}{}
        \\ \color{black!50}\mathsmaller{\mathsmaller{10}} &           & \pilet{s_s = s_L + s_R}{}
        \\ \color{black!50}\mathsmaller{\mathsmaller{11}} &           & \pilet{k_c = kcf\of{k_s, \codestring{SAE}, s_s}}{}
        \\ \color{black!50}\mathsmaller{\mathsmaller{12}} &           & \pilet{k   = pmk\of{k_s, \codestring{SAE}, s_s}}{}
        \\ \color{black!50}\mathsmaller{\mathsmaller{13}} &           & \pinonce{i_L}{} \mbox{\tt{\quad/\!/ send-confirm}}
        \\ \color{black!50}\mathsmaller{\mathsmaller{14}} &           & \pilet{H_L = \cn\of{k_c,i_L,c_L,c_R}}{}
        \\ \color{black!50}\mathsmaller{\mathsmaller{15}} &           & \piwrite{(H_L, i_L)}; \piread{(H_R, i_R)};
        \\ \color{black!50}\mathsmaller{\mathsmaller{16}} &           & \piwrite{\mn{enc}\of{k, m}};
      \end{array}
    \end{equation*}%
  \end{minipage}%
  };
  \coordinate (leftHL)  at (1.8,0.33)   ;
  \coordinate (HLsize)  at (6,5.05)   ;

  \fill[ solid,thick,draw=cyan!70!black,fill=cyan!10!blue!20!white,fill opacity=0.05] (leftHL) rectangle ($ (leftHL) + (HLsize)$);
\end{tikzpicture}
\end{figure}

We highlight the symmetric nature of the protocol by allowing both processes to write their first message to the channel before reading from it.
In other words, the non-deterministic symbolic execution will explore both cases in which either peer will read from the channel first.

The $\pi$-calculus code outside of the box serves to model two aspects:
first, to implement the usage of the same password element $\PE$;
and second, to model the use of the key $k$ after the key exchange to achieve secrecy properties.
In detail, the first line looks for the password element $\PE$ in the table $t$ computed by the participant, either the Leftmost $L$ or the Rightmost $R$, and intended for use with their peer.
If $t$ contains a suitable $\PE$ to communicate to the other party, the participant continues; otherwise, it aborts.
However, it is worth noting that this would not stop an active attacker from attempting to manipulate messages or craft new ones.
Finally, the last line is not part of the scheme and has been artificially added to verify the secrecy of the shared key $k$ through the privacy of the message $m$.
Another modelling choice is to model the counter {\tt send-confirm} with a fresh nonce: using actual numbers would require modelling global state and beyond current ProVerif's capabilities~\cite{cheval2018little}.
This way, we retain tractability and soundness for core cryptographic properties (replay prevention via freshness), at the cost of abstracting away ordering or wrap‑around effects that are implementation‑level concerns rather than cryptographic weaknesses.
The reader may recall the attacks based on cracking counters in the 4-way handshake of WPA2 for key reinstallation~\cite{cremers2020formal}.
In this regard, it is worth noticing that WPA3-SAE, conversely to WPA2, does not implement key reinstallation and does not use counters for key derivation either.
Additionally, {\tt send-confirm} is sent in clear along with a hash that includes it; so, that the adversary knows it regardless of how it is calculated.

\subsubsection*{The two parties}
The SAE protocol illustrated in Fig.~\ref{fig:pi-model-vanilla} is constituted by two participants.
We call one the Leftmost, $L$, and the other the Rightmost, $R$, for the convenience of naming in our model; however, we stress that given its symmetric structure, the protocol can be initiated by either.
All models of the protocol are variations of the same vanilla specification, where we abstract out the pairwise master key as $pmk$ and the EAPOL-Key confirmation key $kcf$.

\subsubsection*{The pre-shared password}
A table $t_p$ of passwords contains all password elements $\PE$ that participants would calculate before engaging the protocol; i.e., $\PE \in \groupset$ is the secret group generator for $L$ and $R$ implemented with finite field cryptography.
From the point of view of the symbolic protocol design, picking a password from a pool and then computing the password element (or generator) is the same as having directly shared the secret password element.

While the protocol can be initited by either party, this is only true for the first phase, since in the second phase, the parties need to know who sends the double and who sends the single hash of the key.
This behaviour is formalised in the key-exchange phase by allowing both parties to output before reading, but it cannot be done in the key confirmation phase.

\subsubsection*{The main process}
Symbolic analysis generally enables security analysis of the parallel execution of protocols.
To this end, in $\pi$-calculus, one can implement the main process, $P$, as the parallel execution of two processes: the Leftmost process $P_L$ and the Rightmost process $P_R$, whose details model the protocol behaviour for each party.
To avoid the unwanted behaviour of flagging false attacks where the initator speaks to itself, a solution is to support the session explicitly between the two parties.
However, the session, denoted as $sID$, is not private and artificially written to the insecure channel $c$, i.e., $\piwrite{sID}$.
Before calling the processes $P_L$ and $P_R$, we establish which party runs $P_L$ and which runs $P_R$.
By doing this, we capture the ability of honest parties to engage with the protocol using either algorithm, whose structure is mirrored in either case.
This choice requires us to ensure that the two parties would not engage in the protocol if they do not share the password element.
For this reason, we also added an environment process $P_P$, which is responsible for inserting shared password elements into a table accessible by $P_L$ and $P_R$ but not by the attacker.

Several security properties rely on custom events being written into the execution trace.
To record such events, we have a special process $P_A$, that records some variables.
In practice, the process $P_A$ would collect information from additional tables containing terms whose contents are ultimately combined, recording a single event of {\em agreement} of terms generated by different entities.
For those security properties that do not require tables to record events, the process $P_A$ will be simply empty, i.e., $\procstop$.

Finally, the main process $P$ that the tool checks has the following structure:
\begin{align*}
 P_{SAE} \gets\ & \pinonce{sID}\ \piwrite{sID}; \\ & \pbr{P_L\of{L,R,sID}\ |\ P_R\of{L,R,sID}} \\
 P \gets\  & (P_P\ |\ P_{SAE}\  |\ P_A)
\end{align*}
that reads informally as the composition of the password setup, the protocol, and the final agreement of terms.

\subsubsection*{Modelling attacks and security properties}
Some attacks or security properties are based on the inspection of protocol execution traces.
Each trace $t$ in the space of all infinite traces $T$ is a potentially endless sequence of generic elements.
A trace is a list that reflects a possible expansion of all combinations of the concurrent executions of a process (a protocol)\footnote{Some tools can reason about an unbounded number of concurrent executions; others are capped.}.
A trace is implicitly temporally ordered, in the sense that $i < j$ implies that $t\at{i}$ (the $i$-th element in the trace $t$) has been recorded before $t\at{j}$ during the expansion.
For some security properties, artificial events must be injected into the process definition.
An event $\traceevent{}$ can be recorded as an element in a trace $t$.
We denote $\traceevent{} \in t$ to say that an event appears in the trace $t$.
Such events can be seen as collections of elements (given as arguments) that the adversary cannot access.
We use them for correctness and authentication.

Attacks and security properties are both statements about execution traces.
While an example is sufficient to demonstrate an attack, conversely, a security property must be a statement (a theorem) over those traces that capture the desirable behaviour of the protocol with global validity.

As introduced in Section~\ref{sec:methodology-security-analysis-protocol}, we modelled the security properties that are desirable in authentication protocols~\cite{hao2018analyzing}:

\paragraph{Correctness} it is captured as the reachability property of the event $\traceevent{Corr}$ at the end of the protocol that includes the exchanged key $k$.
The two processes $P_L$ and $P_R$ would write their entity names, the session and the exchanged key, each in their own table, $t_L$ and $t_R$ respectively, that cannot be read by the adversary, but it is shared among processes.
Then we instantiate the additional process $P_A$ with an event that $\traceevent{Corr}$ that collects information from both tables.
Formally, we first inject the table writing operation
\begin{equation*}
  \piinsert{t_L}{L, R, s, k} \qquad\mbox{ and }\qquad \piinsert{t_R}{R, L, s, k}
\end{equation*}
just after line 7 of the model of $P_L$ described in Fig.~\ref{fig:pi-model-vanilla}.
Then, for all sessions $s$, pairwise master keys $k$, we require that the reasoning core can show a trace $t$ in which the event $\traceevent{Corr}$ is recorded and is such that two honest participants agree on their identities, the password, and the key.
\begin{equation*}
  \forall\; s, k.\; \exists\; t \in T.\; \traceevent{Corr}\of{L, R, s, k, L, R, s, k} \in t.
\end{equation*}
where $L$ and $R$ are the only parties (the adversary is absent\footnote{This can be done in ProVerif with {\tt set attacker = passive.}}), $s$ is the (common) session ID, and $k$ is the pairwise master key.
If such an event is reached, i.e., $\traceevent{Corr}$ is found in $t$, then there exists a run of the protocol in which the two parties have authenticated each other, and they have correctly exchanged the same session key.

\paragraph{Secrecy} we capture the secrect of the exchanged key $k$ and the password element $\PE$ with reachability.
That is, secrecy holds if the adversary cannot reconstruct the secret term by applying all possible combinations of rules to any trace $t$; we denote such (potentially infinite) combinations as $\varrho\of{k_0, t}$, where $k_0$ is the initial knowledge of the adversary.
We capture secrecy of $k$ (or $\PE$) as:
\begin{equation*}
  \forall\; t \in T.\; k \notin \varrho\of{k_0, t} \qquad (\mbox{or }\forall\; t \in T.\; \PE \notin \varrho\of{k_0, t}).
\end{equation*}

\paragraph{Authentication}
we capture weak and strong entity authentication \cite{woo1993semantic}.
Weak entity authentication guarantees that two parties are indeed speaking to each other.
Strong entity authentication requires agreement on other values than the
mere entities.

The standard way to capture authentication in the symbolic model is to verify a correspondence between events across execution traces.
Hence, we inject two events $\traceevent{i}$ in $L$ recording the beginning of the protocol and another event that records its end, $\traceevent{e}$ in $R$.
Informally, $\traceevent{i}$ signifies the belief of the initiator of having started an authentication process with $R$, and $\traceevent{e}$ the belief of $R$ of having terminated an authentication process that must have been started by $L$, i.e., $R$ believes that it is communicating to a genuine $L$.
In particular, if for all the traces $t$ of the symbolic execution of $P$ (defined in Section~\ref{sec:pi-models}), the presence of the {\em final} event $\traceevent{f}$ in $t$ is always after a single {\em initial} corresponding presence of $\traceevent{i}$ in the same trace, then we have verified authentication.
For mutual authentication, we need to inject other two analogous events inverting the participants.
This correspondence of events can be formally described as
\begin{equation*}
  \forall t \in T.\ t\at{j} = \traceevent{e} => \exists i < j.\ t\at{i} = \traceevent{i}.
\end{equation*}
We simplify this notation by removing indices.
We can write equivalently $\forall t \in T.\ \traceevent{e} \in t => \traceevent{i} \in t \land \traceevent{i} < \traceevent{e}$.

To capture weak authentication, we record events with arguments $L$ and $R$
\begin{equation*}
  \forall t \in T.\ \traceevent{e}\of{L,R} \in t => \traceevent{i}\of{L,R} \in t \land \traceevent{i}\of{L,R} < \traceevent{e}\of{L,R}.
\end{equation*}
Stronger notion of authentication adds additional elements as arguments, i.e., the session $s$ and the master kepy pair $k$:
\begin{align*}
  \forall t \in T, s, k.\ & \traceevent{e}\of{L,R,s,k} \in t => \traceevent{i}\of{L,R,s,k} \in t \land \\
 & \land \traceevent{i}\of{L,R,s,k} < \traceevent{e}\of{L,R,s,k}.
\end{align*}

\paragraph{Perfect forward secrecy} this property means that if password elements are compromised, the past keys derived from such passwords still remain secret.
We can reformulate this concept as a passive adversary whom is given one valid password element, and is allowed to eavesdrops (an unbounded number of) executions of the protocol trying to reconstruct any of the keys.

Formally, if we parametrise the protocol with the message, $\pi[m]$, we say that the protocol $\pi$ entails {\em non-interference} of $m$ if and only if
\begin{equation*}
  \forall\; m', m''.\; \pi_m \obseq \pi_{}
\end{equation*}
where $\obseq$ denotes {\em observational equivalence}, i.e., changing the secret $m$ does not change the attacker's observable behaviour.
This is called non-interference and in ProVerif can be modelled by leaking $\PE$ to a passive adversary, then querying for the property {\tt noninterf m}.
Intuitively, if $m$ remains secret, it trivially follows that $k$ is at least as secret as $m$.
Indeed, since the decryption function is public, the reconstruction of $k$ would irredeemably compromise the secrecy of $m$.

\subsubsection*{Handling Unsupported Mathematical Operations}
\label{sec:division}
The reader who is more acquainted with the $\pi$-calculus must have already noticed that the mathematical theory behind the operation $K = \left( \PE^{s_R}E_R \right)^{r_L}$ in Figure~\ref{fig:pi-model-vanilla} is not supported by ProVerif.
ProVerif only supports the exponential operation, whose commutativity can be defined as described in its manual when reasoning about the Diffie-Hellman Key Exchange protocol.
Conversely, multiplications (and additions) can only be partly defined.
They cannot be fully supported for the simple reason that commutativity and associativity would lead the reasoning core to infinite expansions, thus to non-termination.
If the protocol allows it, the designer can model only part of them: we model commutativity but do not need associativity.
This choice is not unusual; for example, we reference the $\pi$-calculus models of recent TLS~\cite{bhargavan2017verified}.

An additional complication occurred in the process of defining the division as the inverse operation to the multiplication: modelling the division is {\em required}, otherwise the $K$ cannot be reconstructed by both peers and {\em correctness} and {\em secrecy} cannot be reasoned about.
To overcome the above known limitation, we modelled the division modulo $p$, denoted as ${\div_p}$, as an {\em extended destructor}.
Such destructors are defined by subexpressions; then, ``each subexpression [...] is rewritten by trying the rewrite rules of [the destructor] in the order given [...], and applying the first applicable rewrite rule.''~\cite{blanchet2022proverif}.

In detail, we first tried to implement $\div_p$ as a function and an equation; however, ProVerif could not handle its equations, as it could not prove their convergence (thus termination).
Our next step was then to add their convergence axiomatically; alas, ProVerif does not allow adding such axiomatic behaviour (and skip termination check) if more than one equation uses the same function; in our case, the {\em blocking} function is the standard and (the only) documented equation to be used to model group exponentiation from the ProVerif manual.

Finally, we were successful with the following {\em extended destructor}:
\begin{align*}
  \div_p \equiv\ & \left[ \forall a \in \groupset, x, y \in \natset.\ a^{x+y} \div_p a^y = a^x, \right. \\
   & \forall a,b \in \groupset.\ a \div_p b^{-1} = ab, \\
   & \left. \forall a,b \in \groupset.\ ab \div_p b = a \right].
\end{align*}
Clearly, those rewriting rules cannot define the division comprehensively, but are reducing from left to right (to reach termination) and are enough to reconstruct the final key.
They are even slightly more expressive than strictly necessary to reconstruct the key; this enriches the adversarial capability and allows it to enjoy (marginally) stronger guarantees.

\subsection{Symbolic Models of Device Level}
\label{sec:asmeta-models}

After a traditional description of the SAE protocol, the IEEE 802.11 standard adds a subsection that instantiates the protocol using a finite-state machine.
We formalized this state machine using Abstract State Machines (ASMs), implemented in the ASMETA framework~\cite{asmeta}. 
As illustrated in Fig.~\ref{fig:sae-device}, the device-level model encompasses three distinct agents: the Service Management Entity (SME), the Parent Process (PP), and Protocol Instances (PIs).

Our ASM model implements the complete device-level SAE protocol through three interacting agents, each modelled as a separate ASM specification.

The SME is the highest-level management component that initiates and terminates SAE exchanges.
It generates two primary events:
\begin{itemize}
  \item \texttt{INITIATE}: Signals the PP to begin SAE with a designated peer identified by MAC address
  \item \texttt{KILL}: Instructs the PP to remove all protocol instances associated with a peer's MAC address
\end{itemize}

The SME operates independently of the protocol mechanics and serves as the interface between higher-level network management and the SAE protocol execution.

The PP maintains the global state of all active SAE exchanges and manages the lifecycle of Protocol Instances.
Its responsibilities include:
\begin{itemize}
    \item Receiving and dispatching events from the SME
    \item Allocating new PIs in response to \texttt{INITIATE} events
    \item Maintaining the \texttt{Open} counter tracking active instances
    \item Routing incoming SAE frames to appropriate PIs
    \item Enforcing the anti-clogging token mechanism when \texttt{Open} exceeds the threshold
    \item Managing the database of active PIs indexed by peer MAC addresses
\end{itemize}

The PP ensures that for any given peer identity, there exists at most one PI in \texttt{Committed} or \texttt{Confirmed} state, a critical invariant for protocol correctness.

Each PI represents a single SAE exchange with a specific peer. A PI transitions through five states as shown in Fig.~\ref{fig:sae-device}:
\begin{itemize}
    \item \texttt{Nothing}: Initial state after creation
    \item \texttt{Committed}: After sending SAE Commit message
    \item \texttt{Confirmed}: After sending SAE Confirm message
    \item \texttt{Accepted}: After successful mutual authentication
\end{itemize}
Multiple PIs can exist simultaneously for different peers, but the PP enforces strict constraints on PI multiplicity per peer to prevent certain classes of attacks, e.g., resource drain~\cite{vanhoef-sp2020-dragonblood}.

\subsubsection{Model Refinement Through Scenario-Based Validation}
\label{subsec:model-refinement}

Our ASM model underwent extensive refinement through scenario-based testing using Avalla~\cite{carioni2008scenario}, the scenario validation tool of the ASMETA framework. This iterative process revealed numerous inconsistencies and ambiguities in the IEEE 802.11 standard specification that were not apparent from the graphical state machine representation.

Fig.~\ref{fig:avalla-sim} shows a representative scenario testing the handling of a \texttt{COM} event when a PI is in the \texttt{ACCEPTED} state.
\begin{figure}[htbp!]
\caption{Scenario-based testing of the \texttt{COM} event when PI is in \texttt{ACCEPTED} state (simplified, showing first step of three).}
\label{fig:avalla-sim}
\begin{lstlisting}[frame=lines,language=AsmetaL,basicstyle=\smaller\sffamily]
scenario PI_Accepted_SYNC
load SAE_20_scenari.asm
...
set select_mac_sme_signal := MAC1;
set signal := NOSIGNAL;
set select_mac := MAC2;
set message_from_peer(MAC2) := EMPTY;

exec extend ProtocolInstance with $pi do
  par
    protocol_Instance_Association($pi):=MAC1
    database_instance($pi):=true
    state_MAC($pi):=ACCEPTED
    database_instance_counter(MAC1):=1
    save_PI:=asSequence(ProtocolInstance)
    start_Timer($pi,TIMER0):=false
    sync($pi):=6
    sc($pi):=0
  endpar;
step
...
\end{lstlisting}
\end{figure}
The refinement process identified multiple error categories:
(i) \textbf{missing events}: several state transitions lacked specification of events (e.g., \texttt{COM} event not specified when passing frames);
(ii) \textbf{incorrect event names}: the specification used \texttt{Sync} (a variable) instead of \texttt{big(Sync)} (the event triggered when the sync counter exceeds the threshold);
(iii) \textbf{undefined behaviours}: multiple error conditions (e.g., password identifier mismatch, commit validation failure) did not specify how to be handled;
(iv) \textbf{missing state transitions}: error paths lacked \texttt{Del} event generation; and
(v) \textbf{inconsistent conditions}: anti-clogging token checks used inconsistent comparison operators.

The machine-executable semantics of ASMs forced us to make explicit every aspect of the protocol behaviour that the standard left ambiguous.
For example, we discovered the \textbf{Sync Event Error}.
During scenario testing, we attempted to trigger the parent process behaviour when a PI exceeded its sync limit. The scenario repeatedly failed because the rule \texttt{r\_ParentProcess\_Behavior} referenced:
\begin{lstlisting}[language=AsmetaL,basicstyle=\smaller\sffamily]
    if parent_Process_Event($pi) = Sync then ...
\end{lstlisting}
However, no PI rule ever set this event. Examining the PI state machine revealed that \texttt{sync} was a counter variable, not an event. The correct event should have been \texttt{big(Sync)}, triggered when \texttt{sync > dot11RSNASAESync}. This became Erratum 11 in our submitted corrections.

\subsubsection{Formal Verification with Model Checking}
\label{subsec:asm-verification}

After scenario-based refinement, we verified safety properties using AsmetaSMV, which translates ASM specifications into input for NuSMV model checker.
The verification focused on invariants and temporal properties expressed in Computation Tree Logic (CTL).

Our verification effort established several critical safety properties of the device-level SAE protocol through model checking with AsmetaSMV.
The first and most fundamental property concerns the uniqueness of protocol instances per peer.
For any MAC address $m$, the system maintains at most one PI in either \texttt{Committed} or \texttt{Confirmed} state associated with $m$.
This single PI invariant prevents the protocol instance accumulation attacks identified during our analysis, ensuring that concurrent authentication attempts with the same peer cannot lead to resource exhaustion or state confusion.

The second verified property establishes the correctness of the \ttt{Open} counter, which serves as a global resource tracking mechanism for active protocol instances.
We proved that this counter accurately reflects the number of PIs not in the \ttt{Nothing} state at any point during execution. This property is essential for the correct operation of the anti-clogging token mechanism, as the counter value directly determines when the threshold-based protection activates.

Related to counter correctness, we verified the anti-clogging enforcement property, which guarantees that when \ttt{Open} reaches or exceeds
the threshold,
 the parent process correctly rejects all new \ttt{Commit} messages that lack valid anti-clogging tokens.
This property provides formal assurance that the DoS protection mechanism activates as specified in the standard.

The verification also confirmed state reachability within the protocol instance state machine. We established that all states defined in the PI state machine---\texttt{Nothing}, \texttt{Committed}, \texttt{Confirmed}, and \texttt{Accepted}---are reachable from the initial \texttt{Nothing} state under appropriate execution sequences. This property ensures that no states are spuriously defined and that the state machine is properly connected.

Finally, we verified an eventual progression property that eliminates the possibility of infinite loops or deadlocks in protocol execution. Specifically, we proved that any PI currently in \texttt{Committed} or \texttt{Confirmed} state must eventually transition either to the \texttt{Accepted} state (indicating successful authentication) or return to \texttt{Nothing} (indicating termination). This liveness property guarantees that protocol instances do not remain indefinitely in intermediate states, which could otherwise lead to resource leaks or denial-of-service conditions.

\section{Findings and Remediations}
\label{sec:findings}

As introduced in Section~\ref{sec:methodology}, the WPA3-SAE protocol is specified in the traditional way for protocol analysts by explaining how messages are exchanged (Communication level) and the logic behind them, to validate security properties.
In addition, an instance of such a protocol is also described as a state machine that runs on a single device (Device level).
Such additional description focuses on state transitions and interface behaviour, prioritizing properties related to operational correctness within the device's environment.

Table~\ref{tbl:summary-security-properties} reports the classes of checked properties and the verification results.
\begin{table}[htbp]
\caption{Summary of results on formal verification of security properties in ProVerif (left) and ASMETA (right).}
\label{tbl:summary-security-properties}
\begin{center}
\begin{tabular}{lcccccc}
\toprule
& \multicolumn{6}{c}{\textit{Communication level}} \\
\cmidrule{2-7}
\textbf{}             & \textbf{CO} & \textbf{SA} & \textbf{WA} & \textbf{SK} & \textbf{SPE} & \textbf{PFS} \\ 
\midrule
\textbf{IEEE 802.11:2020}    & \rh        & \rc         & \rc         & \rf         & \rf          & \rf          \\
\textbf{Patched models}      & \rf        & \rf         & \rf         & \rf         & \rf          & \rf          \\
\midrule
& \multicolumn{6}{c}{\textit{Device level}} \\
\cmidrule{2-7}
\textbf{}  & & \textbf{SF} & \textbf{DL} & \textbf{ES} & \textbf{RC} &  \\
\midrule
\textbf{IEEE 802.11:2020}  &  & \rc        & \rc         & \rc         & \rc         &  \\
\textbf{Patched models}   &   & \rf        & \rf         & \rf         & \rf         &  \\
\bottomrule
\end{tabular}
\end{center}
\textbf{\em Legend}. Correctness (CO), Strong/Weak authentication (SA)/(WA), Key secrecy (SK), Password element secrecy (SPE), Perfect Forward Secrecy (PFS); Safety (SF), Deadlock (DL), Edge Safety (ES), Reachability (RC).
\textbf{Outcomes}: (\rf) - verified, (\rh) - some cases are verified, (\rc) - attacks found.%
\end{table}%

\subsection{Replay attack}
\label{sec:replay-attack}
The verification of the protocol design of WPA3-SAE in the latest standard~\cite[§12.4]{ieee802.11:2020} highlighted a vulnerability in the logical flow, leading to a break in its authentication in our model.
This was unexpected, as implementations, e.g., \texttt{hostapd}, have been patched for the same in 2019~\cite{vanhoef-sp2020-dragonblood}.

\begin{figure}[htbp]
  \caption{Replay attack at the design level specification of the WPA3-SAE protocol in IEEE 802.11:2020, the variant that uses finite field cryptography.}
  \label{fig:replay-attack}
  \centering
  \newcounter{MsgHeight}
  \setcounter{MsgHeight}{1}
  \newcommand{\coefficient}{0.23}
  \newcommand{\nextmsgbelow}{\stepcounter{MsgHeight}}
  \newcommand{\drawmsg}[4][black]{\draw[-stealth,thick,draw=#1] ($(#2b)!{1-\coefficient*\theMsgHeight}!(#2e)$) -- node[above,midway]{\color{#1}#3}($(#4b)!{1-\coefficient*\theMsgHeight}!(#4e)$);}
  \scalebox{0.75}{%
  \begin{tikzpicture}
    \coordinate (Ab) at (0,0);
    \coordinate (Ae) at (0,3);
    \coordinate (Mb) at (8.3,0);
    \coordinate (Me) at (8.3,3);
    \coordinate (Bb) at (10,0);
    \coordinate (Be) at (10,3);

    \draw[        ,thick] (Ab) -- (Ae) node[pos=1.05]{\ul{Leftmost}};
    \draw[draw=red,thick] (Mb) -- (Me) node[pos=1.05]{\color{red}\ul{Mallory}};
    \draw[        ,thick] (Bb) -- (Be) node[pos=1.05]{\ul{Rightmost}};

    \drawmsg{A}{$s_L, E_L$}{M}
    \drawmsg[gray]{M}{\em blocked!}{B}
    \nextmsgbelow
    \drawmsg{M}{$s_R {\color{gray}\sbr{= s_L}}, E_R \color{gray}\sbr{= E_L}$}{A}
    \nextmsgbelow
    \drawmsg{A}{$\cn\of{k_c,i,s_L,E_L,s_R,E_R} \color{gray}\sbr{= \cn\of{k_c,i,s_L,E_L,s_L,E_L}}$}{M}
    \drawmsg[gray]{M}{\em blocked!}{B}
    \nextmsgbelow
    \drawmsg{M}{$\cn\of{k_c,i,s_R,E_R,s_L,E_L} \color{gray}\sbr{= \cn\of{k_c,i,s_L,E_L,s_L,E_L}}$}{A}
  \end{tikzpicture}}
\end{figure}
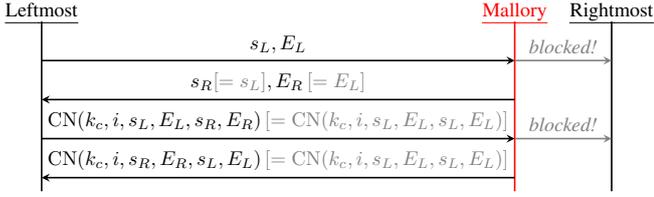

We captured the mechanics of this attack by authenticating $L$ in $R$ (mirroring the process $L$), using the artificial injection of two events, $e_i$ in $P_L$ and $e_f$ in $P_R$.
Strictly following the specification of the SAE protocol, the formal tool ProVerif~\cite{blanchet2022proverif} can reconstruct the flow of an attack as a counterexample of the property described above.
By inspecting the reconstruction, we noticed that it is {\em simply} carried out by blocking commit and confirm messages from $L$ to $R$ when $L$ initiates the protocol and finally reflecting to $L$ its own messages.
Fig.~\ref{fig:replay-attack} shows the mathematical operations for which the replay of messages is (mathematically) acceptable.

The patch for this specific attack would be to discard messages with the same $s_L$, $E_L$ or both.
In our model, we can add the guard ${c_L \neq c_R}$ just after line 7 of the model as $P_L$ described in Fig.~\ref{fig:pi-model-vanilla} (and likewise for $P_R$, which is omitted).
That guard would allow the rest of the model to run and would stop in the case of a replay, i.e., $c_L = c_R$.
With our patch, the same verification no longer produces authentication attacks.

An interesting note is that the protocol instance~\cite[§12.4.8.6.4]{ieee802.11:2020} ignores replayed commit messages (allowing $s_L = s_R$ even when $E_L \neq E_R$ and vice versa), de facto evidencing awareness about this attack.
Remarkably, the reflection attack was known at least as early as February 2015~\cite{rfc7664} (and correctly handled with a guard).
When the 802.11 standard was updated in 2016 to include the protocol, the reflection guard must have been incorrectly removed.
Later on, it was reintroduced incorrectly in the state-machine exemplar instance.
As developers may use paradigms other than state machines, this section is presented as an optional part of the specification -- as it exemplifies an instantiation of the protocol specified in previous sections.
We also remark that the reflection attack is \emph{unnoticeable} by a state-machine analysis, as it does not disrupt any security property typically analysed at that level.
We argue that this is the reason why several implementations suffered from this attack several years after~\cite{vanhoef-sp2020-dragonblood}; Appendix~\ref{sec:real-world} shows how \textsf{hostapd} patched it in 2019.
The patches proposed by \cite{vanhoef-sp2020-dragonblood}, included in the revision under analysis, could not fix the specification for this attack, as their study at the implementation level could not identify the source of the mistake (they could not even refer to the specification).
Conversely, formal analysis makes it clear where it is and how to fix it.

\subsection{Deadlock}
\label{sec:deadlock}
Requirements in the WPA3 SAE standard include safety properties stated in natural language that can be translated into CTL (safety) properties and checked on the ASM model to ensure its robustness and correctness. 
For example, the specification states that: 
\begin{quotes}[§12.4.8.6.1]
  For any given peer identity, there shall be only one protocol instance in the Committed or the Confirmed states.
\end{quotes}
This could be expressed by the following CTL formula stating that, globally ($AG$), there does not exist a state where two PIs ($pi_1$ and $pi_2$) have the same \textit{mac} (i.e., the same identity) but incompatible \textit{state} according to the specifications.
\[
\scalebox{0.8}{%
\begin{minipage}{\linewidth}
\begin{align*}
AG \Bigl( \lnot &  \bigl( \macop\of{pi_1}=\macop\of{pi_2} \land \\
                & \begin{array}[b]{clcl}
                         \bigl( & (\stateop\of{pi_1} = \ms{Committed} &\land & \stateop\of{pi_2} = \ms{Confirmed}) \\
                    \lor  & (\stateop\of{pi_1} = \ms{Confirmed} & \land & \stateop\of{pi_2} = \ms{Committed}) \\
                    \lor  & (\stateop\of{pi_1} = \ms{Confirmed} & \land & \stateop\of{pi_2} = \ms{Confirmed}) \\
                    \lor  & (\stateop\of{pi_1} = \ms{Committed} & \land & \stateop\of{pi_2} = \ms{Committed}) \\
                  \end{array} \bigr) \bigr) \Bigr)
\end{align*}
\end{minipage}
}
\]

Since this property requires checking the internal states of multiple PIs, it was verified against the PP model, which stores the PIs' current state.
The property detects no attacks.

However, the standard documentation does not mention other relevant properties, such as the absence of deadlocks, that are necessary to guarantee the completeness of error-handling cases.
During the analysis of the absence of deadlock in the \textsf{Committed} state, we discovered, through the following failing property, which states that there does not exist a configuration where the PI ($pi$) is in \textit{state} \texttt{Committed} and an error occurs (\textit{fail} holds),
\begin{equation*}
AG \pbr{\lnot EX \pbr{\stateop\of{pi} = \ms{Committed} \land fail}}
\end{equation*}
This shows that the standard does not handle the error when a wrong commit message is received.
This not-handled error causes the PI's state machine to deadlock, with no chance for the PI to be deallocated by the PP.
The standard specifies that a timer \tokens{T0} could have mitigated the deadlock by sending a \textsf{Del} event when it expired.
However, the timer is deactivated when the PI is in the Committed state, and it checks the Commit message's content, preventing the attack mitigation.

The deadlock is even more severe due to the safety property mentioned above because a malicious agent can easily execute the attack.
Indeed, as a result of this deadlock, the PP is unable to create any new instances with the same MAC address as the one associated with the deadlocked PI. 
As a result, this may lead to a Denial-of-Service (DoS) attack on the affected peer, making it unable to connect to the network. 
The attacker could also exploit this vulnerability to cause a peer to run out of memory, as all non-deallocated instances end up in the aforementioned error state. 
The last attack exhibits a pattern similar to that described by \cite{vanhoef-sp2020-dragonblood}, but it exploits a different vulnerability. 
The attack can be solved by adding, in the state machine, two transitions from the \textsf{Committed} state to the \textsf{Nothing} state.
These transitions must involve sending a \textsf{Del} event to the PP to signal it to deallocate the PI.

\subsection{Password Identifiers}
\label{sec:common-findings}

Both analyses at the two levels could find issues due to a novel feature in the 2020 revision of IEEE 802.11~\cite{ieee802.11:2020}: the possibility of using multiple passwords.
This change is meant to enhance security by resisting dictionary attacks, providing (selective) forward secrecy, allowing user flexibility, preventing credential sharing, facilitating secure connections for IoT devices, adapting to evolving threats, and improving overall usability.
To enact this, an SAE entity can {\em require} its peer to use a specific password by sending a password identifier:
\begin{quotes}[§12.4.5.4]
  If the peer's SAE Commit message contains a password identifier, the value of that identifier shall be used in the construction of the password element for this exchange.
\end{quotes}
Failure occurs if no password map exists for that identifier.

\subsubsection{Correctness violation}
We implemented the exchange of password identifiers and their specified behaviour in ProVerif and found that this new feature violate the correctness in some unhandled cases.
On the basis of the code illustrated in Fig.~\ref{fig:pi-model-vanilla}, we modified line 6 of $P_L$ to $\piwrite{\pair{c_L}{p_L}}; \piread{\pair{c_R}{p_R}}$, where $p_L$ and $p_R$ are the requested password identifiers.
An analogous change is made for $P_R$.
We have three cases: if $p_R = \bot$, a password identifier is not requested; if $p_R \neq \bot$ but $p_R$ is not valid, then we abort; otherwise, if $p_R$ is a valid identifier, the password element $\PE'$ depending on $p_R$ is used accordingly.
The following code captures this behaviour:
\begin{equation*}
\begin{array}{llr}
   {p_R \neq \bot}                             & \qquad\  & \mbox{\color{gray}} \\
   \quad\pigetse{t_p}{=\!L, =\!R, =\!p_R, \PE'}{}{}{}  & \mbox{\color{gray}fails if $p_R$ not found} \\
   \quad\pilet{E_L = \PE'^{-m_L}}{}                    \qquad\  & \mbox{\color{gray}recalculate message} \\
   \quad\piwrite{\pair{s_L}{E_L}};                     \qquad\  & \mbox{\color{gray}re-commit}
\end{array}
\end{equation*}
where a missing entry for an invalid $p_R$ in $t_p$ for $L$ and $R$ would automatically fail.
Then, correctness is capture according to its definition in Section~\ref{sec:pi-models}.
We first attempted to write this formalism in a single model.
Unfortunately, ProVerif could not be conclusive ({\tt could not be proven}).
We automated the process in ProVerif by hard-coding all the different cases into separate, independent models.
The only cases left out are those in which the two peers do not share the requested password, as it was trivial in ProVerif to verify that, as expected, the peers could not establish communication.
All cases are illustrated in Table~\ref{tbl:correctness-cases}.
\begin{table}[htbp]
\caption{Expected behaviour of the WPA3-SAE protocol when peers either omit or specify (set) a password identifier corresponding to a pre-shared password.}
\label{tbl:correctness-cases}
\vspace{-3mm}
\begin{center}
\begin{tabular}{rcccc}
\toprule
\textbf{}                & \textbf{omit-omit} & \textbf{set-omit} & \textbf{omit-set} & \textbf{set-set} \\
\midrule
\textbf{same id}      & \rf                & \rf               & \rf               & \rf              \\
\textbf{different id} & \rc                & \rf               & \rf               & \rc              \\
\bottomrule
\end{tabular}
\end{center}
\textbf{\em Legend}. (\rf) - SAE should succeed, (\rc) - SAE should fail.
\end{table}
We remark that the semantics of the \textbf{set-set} case with different identifiers were \textbf{absent in the specification}.
The solution proposed here by us is integrated in the last specification~\cite{ieee802.11:2024} published on 2025.

In our analysis of correctness, if ProVerif is unable to reach $\traceevent{Corr}$, then a key cannot be exchanged.
In the cases where it should, it violates the security property of correctness.
One such case is when both $P_L$ and $P_R$ require a specific password through a valid but different password identifier.
In such a case, the password element $\PE$ will be different between the two peers, and they will not be able to verify the peer confirmation messages.
The situation can be remarkably relevant in meshes, where the possibility of two peers initiating simultaneously can be frequent, and about which WPA3-SAE is particularly focused:
\begin{quotes}[§12.4.1]
  SAE shall be implemented on all mesh STAs to facilitate and promote interoperability.
\end{quotes}
In practice, meshes treat all peers uniformly in the network, yet some peers might possess multiple passwords tied to distinct profiles that they can seamlessly switch between.
Again, we notice that the state machine deviates from the protocol specification: two peers starting the protocol with two different password identifiers would fail the authentication.
Even if this {\em seems} just a patch in the wrong place, it is not, and that is why: firstly, it is coincidental, as there is no justification for such a deviation, but more importantly, both peers would act legitimately and see the authentication failing, so they would try again and again.
In other words, there is no mechanism to prevent both peers from repeatedly attempting (and failing) the authentication, leading to obvious non-termination and resource drain.
Additionally, as the map of password identifiers is {\em not bijective}, it could map to the same password with a different index for some reason, allowing the protocol to end correctly -- but with a different password -- leading to impossible communication.
However, we do not model this behaviour.

In Section~\ref{sec:combined-findings}, we will examine how the outlined specifications, supported by formal verification, address the missing specifications.

\subsubsection{Stall on bad password identifiers}
\label{sec:stall-identifiers}
In the standard's documentation, the description of the multiple branches of the state machine that capture the WPA3-SAE protocol behaviour is given in natural language and leaves out relevant details. 
This can lead to ambiguity and misinterpretation, resulting in serious vulnerabilities or attacks during implementation.
For instance, when we modelled the PI state machine with ASMETA, we discovered an attack that was caused by unclear instructions on how the principal PI should handle an error message.
The vulnerability lurks in the protocol requirement describing a case of failure:
\begin{quotes}[§12.4.8.6.3]
  If so, and there is no password associated with that identifier, \textit{BadID} will be set, and the protocol instance will construct and transmit an authentication frame with StatusCode set to \textit{UNKNOWN\_PASSWORD\_IDENTIFIER}
\end{quotes}
However, this requirement infers the following safety property:
$AG \pbr{\lnot \pbr{\stateop\of{pi} = \ms{Nothing} \land \eventop\of{pi} \neq \ms{Del}}}$.
Indeed, in case of a failure, PIs ($pi$) (in the state \textsf{Nothing}) should always request to be deallocated by the PP by sending the termination \textit{event} \textsf{Del}.
Failing, the property returns a trace where the PI remains stuck in the \textsf{Nothing} state.
As shown by the requirement, the  \textsf{Del} event is missing, leading the PI to become unresponsive, which cannot be removed by the PP.

This attack, in turn, leads to the violation of additional safety properties in the PP model.
More precisely, the standard in a note states:
\begin{quotes}[§12.4.8.6.3]
  NOTE—A protocol instance in Nothing state will never receive an SAE Confirm message due to the state machine behaviour of the parent process
\end{quotes}
However, by analysing the behaviour of the PI state machine when it is stalled, we discovered that the PP can send the \tokens{CON} event to the PI that is waiting in the \textsf{Nothing} state (this is because the PP only checks the presence of a PI with the correct MAC but does not check the PI state).
The PI does not handle the \tokens{CON} in the Nothing state, and this could further aggravate the situation, as unhandled severe exceptions could be triggered.

\subsubsection{Patching password identifiers}
\label{sec:combined-findings}

At the Communication level, two peers are not forbidden from choosing different password identifiers.
Still, when the two password identifiers differ, the protocol will fail without exchanging a common key (correctness violation).
Surprisingly, the state machine analysis showed that two peers are {\em not} allowed to choose different password identifiers (failure), contradicting the protocol specification.
We discovered that the only states where a renewal of the commit message could have happened were the $\ms{Commited}$.
The main case of interest turned out to be the agreement on groups, as, if two peers initiated the protocol with different but supported groups, then one group would be selected by both based on the natural ordering of their MAC address: in detail, the one with the lowest MAC has to re-commit and confirm, and the other has to ignore and wait for another commit message.

We leveraged this result back into our model at the Communication level to include group agreement according to the specification, as well as our {\em patch} to handle password identifiers (see Section~\ref{sec:common-findings}). %
From the Communication perspective, we end up finding a {\em false} attack that seems to downgrade password identifiers.

\begin{figure}[h!]
  \caption{Attack on password identifier WPA3-SAE protocol in IEEE 802.11:2020 at the Communication level, that reveals to be a {\em false} attack when we analyse the Device level.}
  \label{fig:false-attack}
  \centering
  \setcounter{MsgHeight}{1}
  \newcommand{\nextmsgbelow}{\stepcounter{MsgHeight}}
  \newcommand{\coefficient}{0.19}
  \newcommand{\drawline}[3][black]{\draw[,thick,draw=#1] ($(#2b)!{1-\coefficient*\theMsgHeight}!(#2e)$) -- node[above,midway]{}($(#3b)!{1-\coefficient*\theMsgHeight}!(#3e)$);}
  \newcommand{\drawmsg}[4][black]{\draw[-stealth,thick,draw=#1] ($(#2b)!{1-\coefficient*\theMsgHeight}!(#2e)$) -- node[above,midway]{\color{#1}#4}($(#3b)!{1-\coefficient*\theMsgHeight}!(#3e)$);}
  \scalebox{0.75}{%
  \begin{tikzpicture}
    \coordinate (Ab) at (0,0);
    \coordinate (Ae) at (0,3);
    \coordinate (Mb) at (7,0);
    \coordinate (Me) at (7,3);
    \coordinate (Bb) at (10,0);
    \coordinate (Be) at (10,3);

    \draw[        ,thick] (Ab) -- (Ae) node[pos=1.05]{\ul{Alice}};
    \draw[draw=red,thick] (Mb) -- (Me) node[pos=1.05]{\color{red}\ul{Mallory}};
    \draw[        ,thick] (Bb) -- (Be) node[pos=1.05]{\ul{Bob}};

    \drawmsg{A}{M}{$mA, \bot, g$}
    \drawmsg{M}{B}{$mA, {\color{red}\bar{p}}, g$}
    \nextmsgbelow
    \drawmsg{B}{A}{$mB, \bar{p}, g'$}
    \nextmsgbelow
    \drawmsg{A}{B}{$mA', \bar{p}, g'$}
    \nextmsgbelow
    \drawmsg{B}{A}{$CN(mB,mA')$}
    \nextmsgbelow
    \drawmsg{A}{B}{$CN(mA',mB)$}
  \end{tikzpicture}}
\end{figure}
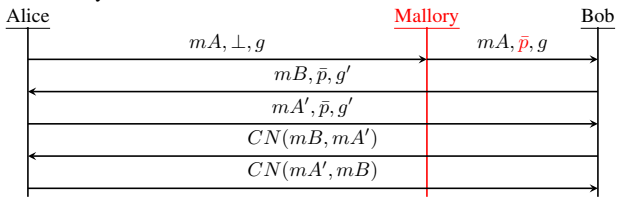

The Communication level model confirms the existence of this attack, allowing two peers to start with different groups and password identifiers.
However, the state machine actually implements a check on this, disallowing the start of the third message in Fig.~\ref{fig:false-attack}.

This shows that addressing the flaw at the Device level without proper justification is masking a vulnerability at the Communication level. This contributes to the significant number of misalignments in our Errata. 
This example illustrates that our combined approach has enabled our formal verification process to achieve a more comprehensive security analysis compared to traditional single-model verification.
We argue that the correct design of the protocol (which our patch accounts for) should supersede the patching of the incorrect design at a different level.
Additionally, the deeper analysis here sketched also suggests that two other important points had to be added to the Communication level specifications: i) despite both peers being allowed to start the protocol, when one receives before having sent its first message, it should consider itself a Receiver and act accordingly; and ii) a Receiver should always accept a supported group, never offer a different group, and terminate the protocol in a failure state, if an unsupported group is offered according to the expected rejection list of groups.
\textit{Our verification shows that with such specifications, the above attack will not occur.}

\section{Conclusions}
\label{sec:conclusions}

We propose the first formalization of WPA3-SAE security protocol in two distinct verification formalisms.
In particular, we combined state-of-the-art modelling techniques for protocol analysis, which are mechanized in ProVerif and in ASMETA.
The IEEE 802.11 specification~\cite{ieee802.11:2020} naturally suggested such a choice, as after a traditional description of the SAE protocol (§12.4.1-7), it proceeds to an instantiation at a single-device level described as a finite automaton (§12.4.8).
Our analysis identified several sources of ambiguity, inconsistencies, and contradictions, as well as missing specifications and mistakes, which could both explain some known attacks and detect further potential issues.
Our results include security issues and strongly support the patches that we proposed.
We reported them in the form of over 20 errata to the IEEE 802.11 Working Group during formal discussions, including Dan Harkins, inventor of the SAE protocol under analysis, as well as developers and engineers of several network device manufacturers.
Most of our patches have been vetted and accepted, and are now integrated to the latest revision of the IEEE 802.11 Standard~\cite{ieee802.11:2024}.

\section*{Acknowledgment}
We gratefully acknowledge Dan Harkins and the IEEE 802.11 Working Group for their constructive engagement and valuable technical discussions regarding our proposed patches to the SAE protocol specification. We were invited, as designated experts, to present and discuss our contributions on two occasions. These exchanges were conducted in a collaborative spirit, with a shared objective of strengthening the security of the specification while minimising ambiguity and interpretability. Their openness to scrutiny and commitment to improving the robustness of the standard are sincerely appreciated.

We also thank D. Basin for constructive comments and suggestions, which helped improve the clarity and quality of this work.

\bibliographystyle{IEEEtran}
\bibliography{references}

\newpage
\appendices

\section{A quick but curious look at the hostapd implementation}
\label{sec:real-world}

We refer to the latest version of \textsf{hostapd} (\url{https://w1.fi/} - \textsf{hostapd} 2.10), a widely used open-source network driver that implements WPA3-SAE.
We investigate its adherence to the standard by manually inspecting its source code.
We remark that its latest version already implements all patches to vulnerabilities analysed by Vanhoef and Ronen in ~\cite{vanhoef-sp2020-dragonblood}.

\lstset{
    basicstyle=\ttfamily,
    breaklines=true,
    columns=fullflexible,
    frame=single,
    numbers=left,
    numberstyle=\scriptsize, %
    numbersep=5pt, %
    xleftmargin=10pt, %
    framexleftmargin=10pt, %
    keywordstyle=\color{blue},
    commentstyle=\color{green!50!black},
    stringstyle=\color{red!50!black},
    captionpos=b,
    literate={\ \ }{{\ }}1 %
}
As we already mentioned in Section~\ref{sec:replay-attack}, the reflection attack was already known and handled in the Dragonfly IETF version of February 2015.
The reflection attack was incorrectly handled when the protocol was adopted by IEEE 802.11 in 2016, and the mistake has stayed there ever since.
It is no surprise that \textsf{hostapd} (and other implementations) suffered from the reflection attack until 2019~\cite{vanhoef-sp2020-dragonblood} when it was noticed.
The inability of the authors in~\cite{vanhoef-sp2020-dragonblood} to correctly suggest patches to the Standard related to the reflection attack adds significance to our approach, showing how formal methods allow for a significant complementary analysis to direct tests on implementations.

A notable deviation from the standard can be seen during group negotiations. The hostapd implementation mishandles the negotiation by causing communication to hang if the parties involved propose two valid but differing groups. Instead of successfully negotiating a valid group~\cite[§12.4.8.6.4]{ieee802.11:2020}\footnote{Initiated in the function \texttt{handle\_auth\_sae} in file  \texttt{ieee802\_11.c}  at line 1590.}, the process gets stuck, resulting in resending the Commit message, which is never accepted. 
Furthermore, the process is not deallocated even if the peer matches the groups but introduces an error in the Commit message (e.g. scalar out of the range $1<scalr<r$), leading to a waste of resources and potential deadlock as per Section~\ref{sec:deadlock} (Errata 16, Artefact).
A similar scenario arises while handling the \texttt{WPA\_EVENT\_SAE\_UNKNOWN\_PASSWORD\_IDENTIFIER}.
The common denominator for these errors is the unclear or missing wording in the specification regarding the use of Del and Fail events, as well as whether a response is sent or if the PI should be deallocated.
Unsurprisingly, as shown by our formal model in Section~\ref{sec:stall-identifiers}, the error handling in the implementation of an unknown password identifier produces the same vulnerability and does not de-allocate the PI, leading to a stall with a consequent waste of resources.
An excerpt of the code that leads to the weakness is shown below with the problem visible and attributable to the use of the \texttt{goto} at line 9 that prevents the execution of the instance removal at line 10.

\begin{lstlisting}[frame=lines,language=C,basicstyle=\smaller\sffamily,showstringspaces=false]
if (resp == WLAN_STATUS_UNKNOWN_PASSWORD_IDENTIFIER) {
    wpa_msg(hapd->msg_ctx, MSG_INFO,
        WPA_EVENT_SAE_UNKNOWN_PASSWORD_IDENTIFIER
        MACSTR, MAC2STR(sta->addr));
    sae_clear_retransmit_timer(hapd, sta);
    sae_set_state(sta, SAE_NOTHING,
              'Unknown Password Identifier');
    if (sta->sae->state == SAE_NOTHING)
        goto reply;
    goto remove_sta;
\end{lstlisting}

Our analysis reveals significant deviations from the WPA3 standard in the most widely used open-source implementation, posing serious security risks.
We also have obtained strong evidence of our findings, validating them with weaknesses in the code and highlighting the importance of inputs from formal analysis.

\section{IEEE 802.11 SAE Errata Corrections}
\label{app:sae-errata}

This appendix summarizes the errata corrections accepted for the Simultaneous Authentication of Equals (SAE) protocol in IEEE Std 802.11-2020. These corrections address ambiguities, missing specifications, and potential security vulnerabilities identified through formal verification analysis.

\subsection{Password Identifier Handling}
\label{subsec:password-id}

\textbf{Location:} §12.4.3 (Representation of a Password)

\textbf{Issue:} The standard was unclear about behavior when peers exchange different password identifiers, and did not specify restrictions for mesh STAs.

\textbf{Resolution:} Modified §12.4.3 to explicitly state that mesh STAs shall not use password identifiers, addressing potential deadlock situations and DoS attacks that could arise from conflicting identifier exchanges.

\textbf{Rationale:} Formal verification with ProVerif demonstrated that underspecification could lead to DoS attacks when both parties send different password identifiers.

\subsection{Principal Role Clarification}
\label{subsec:principal-role}

\textbf{Location:} §12.4.2 (Assumptions on SAE)

\textbf{Issue:} Inconsistent terminology regarding initiator/responder roles conflicted with SAE's role-less design philosophy.

\textbf{Resolution:} Revised §12.4.2 to clarify that SAE does not force peers into strict roles, while acknowledging that each side may view itself as the initiator for a particular protocol run.

\textbf{Discussion:} The revision maintains consistency with SAE's simultaneous initiation capability while removing contradictory references to strict roles.

\subsection{Anti-Clogging Token Threshold}
\label{subsec:anti-clogging}

\textbf{Location:} §12.4.6 (Anti-Clogging Token mechanism), page 2471

\textbf{Issue:} Inconsistent use of comparison operators (``greater than'' vs. ``greater than or equal to'') when checking the Open variable against the anti-clogging threshold.

\textbf{Resolution:} Changed the condition to ``greater than or equal to'' throughout to ensure consistent token requirement enforcement.

\textbf{Rationale:} The protocol requires token usage when Open equals the threshold, but the edge case was not properly handled, potentially allowing protocol instances with valid tokens to be incorrectly rejected.

\subsection{Protocol Instance Indicators}
\label{subsec:indicators}

\textbf{Location:} §12.4.8.5.2 (Protocol instance variables)

\textbf{Issue:} The purpose of the six protocol instance indicators was not clearly documented.

\textbf{Resolution:} Added explanatory text stating that indicators describe the inner workings of the state machine and are useful for debugging purposes.

\textbf{Discussion:} Indicators track reasons for specific state machine paths and error conditions, improving implementation clarity.

\subsection{SME Event Parameters}
\label{subsec:sme-events}

\textbf{Location:} §12.4.8.3.1 (Parent process events) and §12.4.8.6.1 (Parent process behavior)

\textbf{Issue:} Confusing and inconsistent references to event parameters, particularly mixing ``Init'' and ``Initiate'' events.

\textbf{Resolution:} 
\begin{itemize}
\item Clarified that Initiate and Kill events identify peers by MAC address
\item Corrected references from ``Init event'' to ``Initiate event'' in behavioral descriptions
\item Added explicit MAC address parameter documentation
\end{itemize}

\subsection{Sync Event Correction}
\label{subsec:sync-event}

\textbf{Location:} §12.4.8.6.1 (Parent process behavior)

\textbf{Issue:} Referenced non-existent ``Sync'' event instead of the correct ``big(Sync)'' event.

\textbf{Resolution:} Changed ``Sync'' to ``big(Sync)'' to match the actual event triggered when the Sync counter exceeds its maximum threshold.

\textbf{Rationale:} Sync is a variable, not an event; the associated event is big(Sync).

\subsection{Fail Event Generation}
\label{subsec:fail-event}

\textbf{Location:} §12.4.8.6.5 (Protocol Instance behavior—Confirmed state) and Figure 12-4

\textbf{Issue:} The Fail event was defined but never generated in the state machine, despite being distinct from Del events for password mismatch scenarios.

\textbf{Resolution:} 
\begin{itemize}
\item Added Fail event generation when BadAuth indicator is set (authentication failure in Confirmed state)
\item Updated state transition diagram: (Con, BadAuth, !big(sync)/Fail
\item Modified §12.4.8.6.5 to send Fail event when SAE Confirm message verification fails
\end{itemize}

\textbf{Discussion:} Distinguishing Fail (password mismatch) from Del (other errors) provides better error diagnostics.

\subsection{Auth Event Generation}
\label{subsec:auth-event}

\textbf{Location:} §12.4.8.6.5 (Protocol Instance behavior—Confirmed state)

\textbf{Issue:} Auth event was defined but never sent upon successful authentication.

\textbf{Resolution:} Added Auth event transmission to parent process when protocol instance transitions to Accepted state after successful SAE Confirm message verification.

\subsection{COM Event for Commit Messages}
\label{subsec:com-events}

\textbf{Locations:} §12.4.8.6.1 (Parent process behavior), two instances

\textbf{Issue:} State machine requires events to trigger transitions, but passing of SAE Commit frames did not specify the associated Com event.

\textbf{Resolution:}
\begin{itemize}
\item When passing frame to existing protocol instance: explicitly stated ``as a Com event''
\item When creating new protocol instance with valid anti-clogging token: added ``pass the frame to it as a Com event''
\end{itemize}

\textbf{Rationale:} Protocol instance state machines require explicit events; without specifying Com event, implementations would lack clear guidance.

\subsection{Password Identifier Error Handling}
\label{subsec:password-id-error}

\textbf{Location:} §12.4.8.6.3 (Protocol instance behavior—Nothing state)

\textbf{Issue:} Missing state transition and event after sending UNKNOWN\_PASSWORD\_IDENTIFIER status code.

\textbf{Resolution:} Added Del event transmission to parent process when unknown password identifier is encountered, preventing protocol instance accumulation.

\textbf{Rationale:} Without proper cleanup, attackers could create multiple instances of the same peer identity, violating the single-instance principle and enabling DoS attacks.

\subsection{Silent Discard on Commit Validation Failure}
\label{subsec:commit-validation}

\textbf{Location:} §12.4.8.6.3 (Protocol instance behavior—Nothing state)

\textbf{Issue:} Unspecified behavior when SAE Commit message validation fails (unsupported groups, invalid scalar/element ranges, identity element).

\textbf{Resolution:} Added explicit instruction to ``silently discard the frame'' when sending Del event to parent process.

\textbf{Rationale:} Poor error message handling has historically led to attacks; explicit guidance prevents implementations from making insecure choices.

\subsection{Algorithm Identifier Terminology}
\label{subsec:algorithm-id}

\textbf{Location:} §12.4.8.6.4 (Protocol instance behavior—Committed state)

\textbf{Issue:} Term ``algorithm identifier'' was never defined or used elsewhere in the specification.

\textbf{Resolution:} Revised text to clearly indicate rejection of the offered group in the SAE Commit message with UNSUPPORTED\_FINITE\_CYCLIC\_GROUP status code.

\subsection{Group Mismatch Error Handling}
\label{subsec:group-mismatch}

\textbf{Location:} §12.4.8.6.4 (Protocol instance behavior—Committed state)

\textbf{Issue:} Missing error handling when SAE Commit processing fails during group negotiation for STA with numerically lesser MAC address.

\textbf{Resolution:} Added Fail event generation and termination when processing fails, otherwise continue with new group selection.

\textbf{Discussion:} Distinguishes between acceptable group renegotiation and processing errors requiring termination.

\subsection{Rejection Frame Terminology}
\label{subsec:rejection-frames}

\textbf{Location:} §12.4.8.6.4 and §12.4.8.6.5 (Committed and Confirmed states)

\textbf{Issue:} Undefined term ``rejection frames'' and incorrect treatment of SAE\_HASH\_TO\_ELEMENT status as error.

\textbf{Resolution:}
\begin{itemize}
\item Replaced with specific condition: ``Status is a nonzero value other than SAE\_HASH\_TO\_ELEMENT''
\item Correctly treats SAE\_HASH\_TO\_ELEMENT as valid status rather than error
\end{itemize}

\subsection{Deterministic PWE Generation}
\label{subsec:deterministic-pwe}

\textbf{Location:} §12.4.4.3.2 (Generation of password element with FFC groups by looping)

\textbf{Issue:} Described hunt-and-peck procedure as ``random'' when it is actually deterministic.

\textbf{Resolution:} Removed the word ``random'' from the description.

\subsection{PT Identity Element Handling}
\label{subsec:pt-identity}

\textbf{Location:} §12.4.4.3.3 (Direct generation of password element with FFC groups)

\textbf{Issue:} Algorithm stated probability of PT = 1 should be neglected without providing handling procedure.

\textbf{Resolution:} Added iterative algorithm that appends constant message to label and regenerates PT if PT = 1 occurs, ensuring PT is never the identity element.

\textbf{Discussion:} While negligible with defined groups, algorithmic completeness requires handling this edge case.

\subsection{Replay Attack Prevention}
\label{subsec:replay-prevention}

\textbf{Location:} §12.4.5.4 (Processing of peer's SAE Commit message)

\textbf{Issue:} Scalar validation did not check for replayed commit messages (receiving own scalar back).

\textbf{Resolution:} Added explicit check that processing fails if peer's scalar or element are identical to those sent by the SAE finite state machine, when state is other than Nothing.

\textbf{Rationale:} Protects against reflection attacks where an attacker replays the victim's own commit message.

\subsection{Error Handling Order}
\label{subsec:error-order}

\textbf{Location:} §12.4.8.6.3 (Protocol instance behavior—Nothing state)

\textbf{Issue:} Behavioral specification was unclear about order of operations for error handling.

\textbf{Resolution:} Changed to ``shall silently discard the frame and send a Del event to the parent process'' to match language patterns elsewhere in specification.

\subsection{Summary}

Nineteen errata were accepted (through eighteen changes), some with revisions, and they address:
\begin{itemize}
\item \textbf{Security vulnerabilities:} DoS attacks, replay attacks, reflection attacks
\item \textbf{State machine ambiguities:} Missing events, unclear transitions, undefined behaviors
\item \textbf{Implementation guidance:} Error handling, event generation, status code interpretation
\item \textbf{Terminology consistency:} Standardized language across specification
\item \textbf{Edge case handling:} Anti-clogging tokens, group negotiation, password identifiers
\end{itemize}

These corrections significantly improve the security, implementability, and clarity of the SAE protocol specification, with many issues identified through formal verification techniques using ProVerif.
 
\end{document}